\documentstyle[12pt,emlines]{article}
\def\d{\partial}

\def\bea{\begin{eqnarray}}
\def\eea{\end{eqnarray}}

\def\beq{\begin{equation}}
\def\eeq{\end{equation}}
\def\ba{\beq\new\begin{array}{c}}
\def\ea{\end{array}\eeq}
\def\be{\ba}
\def\ee{\ea}
\def\stackreb#1#2{\mathrel{\mathop{#2}\limits_{#1}}}

\makeatletter
\newdimen\normalarrayskip              
\newdimen\minarrayskip                 
\normalarrayskip\baselineskip
\minarrayskip\jot
\newif\ifold             \oldtrue            \def\new{\oldfalse}
\def\arraymode{\ifold\relax\else\displaystyle\fi} 
\def\eqnumphantom{\phantom{(\theequation)}}     
\def\@arrayskip{\ifold\baselineskip\z@\lineskip\z@
     \else
     \baselineskip\minarrayskip\lineskip2\minarrayskip\fi}
\def\@arrayclassz{\ifcase \@lastchclass \@acolampacol \or
\@ampacol \or \or \or \@addamp \or
   \@acolampacol \or \@firstampfalse \@acol \fi
\edef\@preamble{\@preamble
  \ifcase \@chnum
     \hfil$\relax\arraymode\@sharp$\hfil
     \or $\relax\arraymode\@sharp$\hfil
     \or \hfil$\relax\arraymode\@sharp$\fi}}
\def\@array[#1]#2{\setbox\@arstrutbox=\hbox{\vrule
     height\arraystretch \ht\strutbox
     depth\arraystretch \dp\strutbox
     width\z@}\@mkpream{#2}\edef\@preamble{\halign
\noexpand\@halignto
\bgroup \tabskip\z@ \@arstrut \@preamble \tabskip\z@ \cr}%
\let\@startpbox\@@startpbox \let\@endpbox\@@endpbox
  \if #1t\vtop \else \if#1b\vbox \else \vcenter \fi\fi
  \bgroup \let\par\relax
  \let\@sharp##\let\protect\relax
  \@arrayskip\@preamble}
\def\eqnarray{\stepcounter{equation}%
              \let\@currentlabel=\theequation
              \global\@eqnswtrue
              \global\@eqcnt\z@
              \tabskip\@centering
              \let\\=\@eqncr
              $$%
 \halign to \displaywidth\bgroup
    \eqnumphantom\@eqnsel\hskip\@centering
    $\displaystyle \tabskip\z@ {##}$%
    \global\@eqcnt\@ne \hskip 2\arraycolsep
         $\displaystyle\arraymode{##}$\hfil
    \global\@eqcnt\tw@ \hskip 2\arraycolsep
         $\displaystyle\tabskip\z@{##}$\hfil
         \tabskip\@centering
    &{##}\tabskip\z@\cr}
\begingroup\ifx\undefined\newsymbol \else\def\input#1 {\endgroup}\fi
\newfont{\hr}{msbm10}
\newfont{\ams}{msam10}

\font\numbers=cmss12
\font\upright=cmu10 scaled\magstep1
\def\stroke{\vrule height8pt width0.4pt depth-0.1pt}
\def\topfleck{\vrule height8pt width0.5pt depth-5.9pt}
\def\botfleck{\vrule height2pt width0.5pt depth0.1pt}
\def\Zmath{\vcenter{\hbox{\numbers\rlap{\rlap{Z}\kern 0.8pt\topfleck}\kern
2.2pt
                   \rlap Z\kern 6pt\botfleck\kern 1pt}}}
\def\Qmath{\vcenter{\hbox{\upright\rlap{\rlap{Q}\kern
                   3.8pt\stroke}\phantom{Q}}}}
\def\Nmath{\vcenter{\hbox{\upright\rlap{I}\kern 1.7pt N}}}
\def\Cmath{\vcenter{\hbox{\upright\rlap{\rlap{C}\kern
                   3.8pt\stroke}\phantom{C}}}}
\def\Rmath{\vcenter{\hbox{\upright\rlap{I}\kern 1.7pt R}}}
\def\Z{\ifmmode\Zmath\else$\Zmath$\fi}
\def\Q{\ifmmode\Qmath\else$\Qmath$\fi}
\def\N{\ifmmode\Nmath\else$\Nmath$\fi}
\def\C{\ifmmode\Cmath\else$\Cmath$\fi}
\def\R{\ifmmode\Rmath\else$\Rmath$\fi}

\newcounter{app}

\def\app{\setcounter{equation}{0}
\def\theequation{\Alph{app}.\arabic{equation}}\par
   \addvspace{4ex}
   \@afterindentfalse
  \secdef\@app\@dapp}

\newcommand\@app{\@startsection {app}{1}{0ex}%
                                   {-3.5ex \@plus -1ex \@minus -.2ex}%
                                   {2.3ex \@plus.2ex}%
                                   {\normalfont\Large\bf}}
\def\@dapp#1{%
{\parindent \z@ \raggedright  \bf #1}\par\nobreak}
\def\l@app#1#2{\ifnum \c@tocdepth >\z@
    \addpenalty\@secpenalty
    \addvspace{1.0em \@plus\p@}%
    \setlength\@tempdima{8em}%
    \begingroup
      \parindent \z@ \rightskip \@pnumwidth
      \parfillskip -\@pnumwidth
      \leavevmode \bfseries
      \advance\leftskip\@tempdima
      \hskip -\leftskip
      #1\nobreak\hfil \nobreak\hb@xt@\@pnumwidth{\hss #2}\par
    \endgroup\fi}
\newcounter{sapp}[app]

\def\sapp{\def\theequation{\Alph{app}.\arabic{equation}}
\par
\@afterindentfalse
  \secdef\@sapp\@dsapp}
\newcommand{\@sapp}{\@startsection{sapp}{2}{\z@}%
                                     {-3.25ex\@plus -1ex \@minus -.2ex}%
                                     {1.5ex \@plus .2ex}%
                                     {\normalfont\large\bfseries}}

\def\@dsapp#1{%
{\parindent \z@ \raggedright  \bf #1
}\par\nobreak}
\newcommand{\l@sapp}{\@dottedtocline{2}{1.5em}{2.3em}}

\def\stackreb#1#2{\mathrel{\mathop{#2}\limits_{#1}}}

\def\d{\partial}

\def\2{{1\over 2}}
\def\N2{${\cal N}=2$}

\def\be{ \begin{eqnarray} }
\def\ee{ \end{eqnarray} }

\def\d{\partial}

\def\bea{\begin{eqnarray}}
\def\eea{\end{eqnarray}}

\def\beq{\begin{equation}}
\def\eeq{\end{equation}}
\def\ba{\beq\new\begin{array}{c}}
\def\ea{\end{array}\eeq}
\def\be{\ba}
\def\ee{\ea}
\def\stackreb#1#2{\mathrel{\mathop{#2}\limits_{#1}}}

\begin{document}
\begin{flushright}
ITEP/TH-53/98\\
FIAN/TD-05/98\\
hepth/9812078
\end{flushright}
\vspace{0.5cm}
\begin{center}
{\LARGE \bf Seiberg-Witten theory for a non-trivial
compactification from five to four dimensions}
\vspace{0.5cm}

\setcounter{footnote}{1}
\def\thefootnote{\fnsymbol{footnote}}
{\Large H.W.Braden\footnote{Department of Mathematics and Statistics,
University of Edinburgh, Edinburgh EH9 3JZ Scotland;
e-mail address: hwb@ed.ac.uk },
A.Marshakov\footnote{Theory
Department, Lebedev Physics Institute, Moscow
~117924, Russia; e-mail address: mars@lpi.ac.ru}\footnote{ITEP,
Moscow 117259, Russia; e-mail address:
andrei@heron.itep.ru},
A.Mironov\footnote{Theory
Department, Lebedev Physics Institute, Moscow
~117924, Russia; e-mail address: mironov@lpi.ac.ru}\footnote{ITEP,
Moscow 117259, Russia; e-mail address:
mironov@itep.ru}, A.Morozov\footnote{ITEP, Moscow
117259, Russia; e-mail address: morozov@vx.itep.ru}
}\\
\end{center}
\bigskip
\begin{quotation}
The prepotential and spectral curve are described for a smooth
interpolation between an  enlarged $N=4$ SUSY and ordinary $N=2$ SUSY
Yang-Mills theory in four dimensions, obtained by compactification from five
dimensions
with non-trivial (periodic and antiperiodic) boundary conditions.
This system provides a new solution to the generalized WDVV equations.
We show that this exhausts all possible solutions of a given functional
form.
\end{quotation}
\setcounter{footnote}{0}
\setcounter{equation}{0}

\paragraph{1.}

In their pioneering paper \cite{SW}, N.Seiberg and E.Witten  suggested that
a hidden (duality) symmetry of the Yang-Mills dynamics could be exploited to
provide an ansatz for exact low-energy effective actions for various $N=2$
supersymmetric YM models in four dimensions.
This work uncovered three fundamental objects: a Riemann
surface, the moduli space of the surface and a given meromorphic one-form
on the surface, the Seiberg-Witten
differential $dS$. Such data had also appeared previously in the study of
completely integrable systems and the Seiberg-Witten (SW)
ansatz was interpreted in \cite{GKMMM}
in terms of Whitham dynamics associated with a particular
finite-dimensional integrable system describing the effective dynamics along the
minima in the space of the scalar fields.
The Riemann surface is the spectral curve of the integrable system and the
moduli of the curve are the values taken by the various commuting Hamiltonians.
In the SW context the moduli are the vacuum expectation values of the
various scalar fields and these have also been interpreted as the
 vacuum expectation values fields dynamically develop when a 5-brane
is wrapped around a bare spectral curve \cite{brane}.
Further, the periods $a_i$, $a^{D}_i$ of the Seiberg-Witten
differential are the adiabatic invariants (the action variables $\oint \vec
p d \vec q$) of integrable system and these satisfy the Picard-Fuchs equations
on the moduli space of spectral curves.
These periods define a prepotential  ${\cal F}$ via
$a^D_i = \oint_{B_i}dS\equiv {\d{ F}\over\d a_i}$ and this prepotential
is central to the theory.

Numerous examples of this correspondence between SW-theory and integrable
systems have now been worked out for four
and five-dimensional $N=2$ supersymmetric YM models with various gauge groups
and matter hypermultiplets \cite{int}-\cite{intf}.
Unfortunately this correspondence at present remains phenomenological and
its full explanation is an important outstanding problem in the area.
The purpose of this Letter is to add new and very important model to the list
of known examples. Indeed our example further highlights an additional
feature of the correspondence to which we now turn.

The (perturbative) prepotentials ${\cal F}(a_i)$ of SW-theories
have been observed to satisfy the generalized WDVV equations
on the moduli space \cite{wdvva,wdvvb,wdvvc}:
\be\label{WDVV}
{\cal F}_i{\cal F}_k^{-1}{\cal F}_j =
{\cal F}_j{\cal F}_k^{-1}{\cal F}_i,\ \ \ \ \forall i,j,k;
\ \ \ \ \left({\cal F}_i\right)_{jk}\equiv {\d^3{\cal F}\over
\d a_i\d a_j\d a_k}.
\ee
These equations differ from the standard WDVV equations (of say two-dimensional
topological field theory) which may be cast
${\cal F}_i{\cal F}_1^{-1}{\cal F}_j =
{\cal F}_j{\cal F}_1^{-1}{\cal F}_i$ (for all $i,j$).
Here a particular direction (the $1$)
has been singled out and one further imposes that
${\cal F}_1$ is a constant. Physically this was motivated by the special
status of the vacuum state and two-point correlation functions.
Actually as long as ${\cal F}_k$ is invertible then (\ref{WDVV}) follows from
${\cal F}_i{\cal F}_1^{-1}{\cal F}_j =
{\cal F}_j{\cal F}_1^{-1}{\cal F}_i$ and so (\ref{WDVV}) may be viewed as a
{\it projectivised} form of the usual equations, putting all of the
coordinates on a similar footing. The significant part of the
generalization of \cite{wdvva,wdvvb,wdvvc} lies in not requiring any of the
${\cal F}_k$'s
to necessarily be constant. Indeed the solutions given by the SW-theories are
not.
Again a fundamental understanding of the appearance of (\ref{WDVV}) is
still lacking. From the integrable system side of the correspondence
these equations may be understood in terms of Whitham dynamics \cite{whit}
which enables one to construct a $\tau$-function. Presumably  this corresponds
to  a generating functional for the correlation functions of the light
fields of the theory \cite{RG}.
Although we still await a deeper understanding of (\ref{WDVV}) we may enquire
of its solutions. By doing this one discovers the models of this letter.

We construct our models by first solving the generalized WDVV equations
(\ref{WDVV}) for a general class of perturbative prepotentials
${\cal F}_{pert}$ assuming the functional form
\be\label{funform}
{\cal F}=\sum_{\alpha\in \Phi}f(\alpha\cdot a),
\ee
where the sum is over the root system $\Phi$ of a Lie algebra.
This functional form is motivated by the several existing calculations
that may be found in the literature.
Imposing this ansatz reduces the WDVV equations to a single functional
equation. Such functional equations arise in many guises
in the context of integrable systems \cite{functional} and we are able
to give a general solution, so yielding all solutions of (\ref{WDVV}) of
this given form. Our solution may be interpreted in terms of the
compactification of a five-dimensional  SUSY theory to four dimensions
where supersymmetry breaking is achieved by imposing appropriate boundary
conditions.
The correspondence between
SUSY Yang-Mills theory and an integrable system is then
most
straightforwardly provided by identifying effective charges (couplings)
with period matrices,
\be
{\cal T}_{ij}={\d^2{\cal F}\over\d a_i\d a_j}.
\ee

An outline of the Letter is as follows. In section 2 we begin by
describing a class of models that may arise by compactification of a
five-dimensional  SUSY theory to four dimensions, giving their expected
perturbative prepotential based on known SW solutions.
In section 3 we solve the functional equations described above.
For those less comfortable with functional equations section 4 gives
a physical interpretation of our equation in terms of a boson/fermion
equivalence. Section 5 develops the connection with the spectral curve
and identification of the associated integrable system.
We finish with a brief conclusion.
We shall restrict ourselves in this Letter to the case of $SU(N)$ gauge group.

\paragraph{2.}
Four dimensional $N=2$ supersymmetric YM theories can be obtained by
softly breaking $N=4$ supersymmetric YM in two different ways.
The most commonly discussed way is to directly add a mass term for the
adjoint hypermultiplet in four dimensions and to take the
double-scaling limit $m\to\infty$, $\tau={\theta\over 2\pi}+{4\pi i\over g^2}
\to i\infty$, while keeping $\Lambda^N=m^Ne^{2\pi i\tau}=$ finite.
On the integrable system side this is described in terms of the elliptic
Calogero-Moser model: here $m$ plays the role of the coupling constant and
this double-scaling limit was shown by Inosemtsev to turn the
Calogero-Moser
system into that of the Toda chain. (Various details of this construction
may be found in \cite{intb}.)
An alternative construction of a four dimensional theory is to begin with a
five dimensional $N=2$ supersymmetric YM theory and compactify onto a circle
with periodic boundary conditions. One can  then break $N=4$ SUSY down to $N=2$
by imposing
non-trivial (antiperiodic) boundary conditions on half of the fields.  A smooth
interpolation between the  four dimensional
$N=4$ and $N=2$ models is then provided by the
change of compactification radius from $R=\infty$ to $R=0$.

The perturbative prepotential for  these theories may be calculated.
As explained in detail in \cite{wdvvb}, the perturbative (1-loop)
contribution to the prepotential ${\cal F}_{pert}$ may be
constructed from the mass spectrum and the Coleman-Weinberg type formula:
\be
{\cal F}_{pert}(a)=
\frac{1}{2} \sum_M\sum_{i<j} \pm (a_{ij}+M)^2
\log(a_{ij}+M),\ \ \ a_{ij}\equiv a_i-a_j,\ \ \ \sum a_i=0.
\ee
The choice of sign $\pm$ depends on the nature of the supermultiplet.
In the case of a massive adjoint hypermultiplet (Calogero-Moser
model) we have
\be
\label{ans}
{\cal F}_{pert}^{Cal}(a)=
\frac{1}{2} \sum_{i<j} (a_{ij})^2
\log(a_{ij})
-\frac{1}{2} \sum_{i<j} (a_{ij}+m)^2
\log(a_{ij}+m).
\ee
Suppose now a four dimensional model is obtained by compactification from
five dimensions. Now one will encounter
a whole tower of massive hypermultiplets with $M=2\pi k/R$ and $M=2\pi
(k-1/2)/R$, and
\be\label{pertp}
\begin{array}{rl}
{\cal F}_{pert}^{comp}(a)=&
\frac{1}{2}\left(\frac{ 2\pi}{R }\right)^2\
\sum_{i<j} \sum_{k=-\infty}^\infty \Bigg\lbrace
\left(\frac{Ra_{ij}}{2\pi}+k \right)^2
\log\left(\frac{Ra_{ij}}{2\pi}+k\right)\\
&\quad -\left(\frac{Ra_{ij}}{2\pi}+k-{1\over 2}\right)^2
\log\left(\frac{Ra_{ij}}{2\pi}+k-{1\over 2}\right)
+{\rm regulator}\Bigg\rbrace.
\end{array}
\ee
This will be the form of the prepotential we obtain by seeking solutions to
(\ref{WDVV}).

\paragraph{3.}
Examination of the known perturbative SW prepotentials reveals the structure
(\ref{funform}) as a minimal component. Here we pose the following
general question: for what functions $f$ does the prepotential\footnote{
The case of
compactified five dimensional ($N=1$ SUSY)$\longrightarrow$(4d $N=2$ SUSY)
discussed in \cite{intd,ints}
is not of the type (\ref{ans}), because ${\cal
F}_{pert}(a)$ contains cubic terms depending not only on the
differences $a_{ij}$
\cite{wdvvb}.}
\be\label{pair}
{\cal F}=\sum_{i<j}f(a_{ij}) ,\ \ \ a_{ij}\equiv a_i-a_j,\ \ \ \sum a_i=0
\ee
satisfy the generalized WDVV equations?
Our first result is that (\ref{WDVV}) are satisfied
for the prepotential of the generic form (\ref{pair}) {\em if and only if}
 $g(a)\equiv \left[{\d^3 f\over\d a^3}\right]^{-1}$ satisfies
the functional equation
\be\label{wdvv}
g(a_{12})g(a_{34})-g(a_{13})g(a_{24})+g(a_{14})g(a_{23})=0.
\ee
To show this we extend the analysis of \cite{wdvva}. Two types of entry
appear when evaluating ${\cal F}_i^{-1}{\cal F}_k{\cal F}_j ^{-1}-
{\cal F}_j^{-1}{\cal F}_k{\cal F}_i ^{-1}$: the first corresponds exactly
with (\ref{wdvv}) establishing the necessity, while a second type corresponds
to several combinations of (\ref{wdvv}). Consequently the vanishing
of (\ref{wdvv}) is also a sufficient condition.

Up to the invariance $g(x)\rightarrow \lambda g(\gamma x)$
the general solution of (\ref{wdvv}) is $g(x)=x$, $\sin(x)$, $\sinh(x)$.
This may be argued as follows. Upon setting $a_1=-a_4=x$ and
$a_2=a_3=0$ in (\ref{wdvv}) we deduce that $g(0)=0$. (A priori we do not know
$g(0)$ to be finite.) Similarly upon setting $a_1=a_3=0$, $a_2=-a_4=x$ and
using our first result we find $g(x)$ to be an odd function.
Now apply $\partial_3(\partial_1 +\partial_2)$ to (\ref{wdvv}) and set
$a_1=a_3=a$ and $a_2=a_4=0$. This yields the differential equation
\be
g''(a)g(a)-g'(a)g'(a)+g'(0)g'(0)=0.
\ee
With $\phi(a)\equiv\log g(a)$ this may be reexpressed as the
Liouville type
equation
\be\label{liu}
\phi''(a)=g'(0)^2 e^{-2\phi(a)}
\ee
which has the first integral
$\phi'(a)^2+g'(0)^2 e^{-2\phi(a)}=const\equiv R^2$, and yields
\be
g(a)={g'(0)\over R}\sinh R(a-a_0).
\ee
Now using the oddness of $g(a)$ and the invariance noted above one finds
that $a_0=0$ and
\be\label{g}
g(a)={1\over R}\sinh Ra.
\ee
Corresponding to this solution one finds that
\be\label{dli3}
f(a)=\frac{ {\rm Li}_3 (e\sp{-R a}) -{\rm Li}_3 (-e\sp{-R a})}{2R\sp2},
\ee
where (for $|x|<1$) the trilogarithm is defined by
\be
{\rm Li}_3(x)\equiv \sum_{k=1} {x^{k}\over k^3},
\ee
and this may be extended by analytic continuation \cite{Lewin}.
Indeed, the correspondence with (\ref{pertp}) is most readily done from
the form
\be
g(a)={1\over R}\sin Ra, \qquad f'''(a)=\frac{R}{\sin R a},
\qquad f''(a)=\log\left(\alpha \tan(\frac{R a}{2})\right),
\ee
where $\alpha$ is a constant.
Now upon using the infinite product expansion
\be
f''(a)= \log\left(\alpha\tan (\frac{R a}{2})\right)
      = \log\left(\alpha\frac{Ra}{2}\frac{
             \prod_{k=1}\sp{\infty} (1 - (\frac{ Ra}{2 k \pi})\sp2)
                               }
                               {
             \prod_{k=1}\sp{\infty} (1 - (\frac{ Ra}{2 (k-1/2) \pi})\sp2)
                               }\right)\\
 \ \quad  =\log\left(\alpha\frac{Ra}{2}\prod_{k=1}\sp{\infty}
       \frac{ (\frac{ Ra}{2\pi} +k)(\frac{ Ra}{2\pi} -k)(k-1/2)\sp2}
        {(\frac{ Ra}{2\pi} + k-1/2)(\frac{ Ra}{2\pi} -k+1/2)k\sp2} \right),
\ee
integration yields
\be
f(a)= \frac{1}{2}
a^2 \log\left(\frac{ Ra}{2\pi}\right) +
\left(\frac{ 2\pi}{R }\right)^2\frac{1}{2}\sum_{k=1}\sp{\infty}\bigg\lbrace
\left(\frac{ Ra}{2\pi}+k\right)^2 \log\left(\frac{ Ra}{2\pi}+k\right) \\
\qquad\qquad\qquad
+\left(\frac{ Ra}{2\pi}-k\right)^2 \log\left(\frac{ Ra}{2\pi}-k\right)
-
\left(\frac{ Ra}{2\pi}+k-1/2\right)^2 \log\left(\frac{ Ra}{2\pi}+k-1/2\right)\\
 \qquad\qquad\qquad
- \left(\frac{ Ra}{2\pi}-k+1/2\right)^2 \log\left(\frac{ Ra}{2\pi}-k+1/2\right)
+\alpha_k \left(\frac{ Ra}{2\pi}\right)^2 +\beta_k \left(\frac{ Ra}{2\pi}\right)
+\gamma_k
\bigg\rbrace.
\ee
Here $\alpha_k$, $\beta_k$ and $\gamma_k$ may be chosen to make the sum
convergent. This quadratic term corresponds to the regulator of (\ref{pertp})
and by appropriate regrouping we obtain our earlier prepotential (\ref{pertp}).

We remark that the Calogero-Moser prepotential is known not to satisfy the
WDVV equations
(presumably) because it has an extra modulus $\tau$. However,
in the double-scaling limit (when $g(a)\to a$) it does yield a
solution to the WDVV equations.

\paragraph{4.}
In this paragraph we wish to view our functional equation (\ref{wdvv}) from
a rather different perspective.
This functional equation may be interpreted as the consistency
condition for realising a four-point correlation function in terms of
both bosonic and fermionic operators. To see this, first observe that
(\ref{wdvv}) is equivalent to
\be\label{cor}
H_{1234}\equiv h(a_{12})h(a_{23})h(a_{34})h(a_{41})+
h(a_{13})h(a_{32})h(a_{24})h(a_{41})+
h(a_{14})h(a_{42})h(a_{23})h(a_{31})=0.
\ee
Here $h(a)\equiv {\d^3 f\over\d a^3} =g(a)^{-1}$ and we have used the
oddness of this function.  This is schematically depicted in Fig.1.

\vspace{1cm}

\special{em:linewidth 0.4pt}
\unitlength 1.00mm
\linethickness{0.4pt}
\begin{picture}(155.67,110.67)
\put(11.67,108.00){\circle*{1.89}}
\emline{35.00}{108.00}{1}{11.67}{108.00}{2}
\emline{11.67}{108.00}{3}{11.67}{84.00}{4}
\emline{11.67}{84.00}{5}{35.00}{84.00}{6}
\emline{35.00}{84.00}{7}{35.00}{108.00}{8}
\put(35.00,108.00){\circle*{2.00}}
\put(35.00,84.00){\circle*{2.00}}
\put(12.00,84.00){\circle*{2.00}}
\put(64.00,108.00){\circle*{2.00}}
\put(64.00,84.00){\circle*{2.00}}
\put(88.33,84.00){\circle*{2.00}}
\put(88.33,108.00){\circle*{2.00}}
\emline{64.00}{84.00}{9}{88.33}{108.00}{10}
\emline{88.33}{108.00}{11}{88.33}{84.00}{12}
\emline{88.33}{84.00}{13}{64.00}{108.00}{14}
\emline{64.00}{108.00}{15}{64.00}{84.00}{16}
\put(119.67,84.00){\circle*{2.00}}
\put(119.67,108.00){\circle*{2.00}}
\put(146.33,108.00){\circle*{2.00}}
\put(146.33,84.00){\circle*{2.00}}
\emline{119.67}{83.67}{17}{146.33}{108.00}{18}
\emline{146.33}{108.00}{19}{119.67}{108.00}{20}
\emline{119.67}{108.00}{21}{146.33}{84.00}{22}
\emline{146.33}{84.00}{23}{119.67}{84.00}{24}
\put(105.33,96.00){\makebox(0,0)[cc]{$+$}}
\put(49.00,96.33){\makebox(0,0)[cc]{$+$}}
\put(0.67,95.00){\makebox(0,0)[cc]{$H_{1234}=$}}
\put(155.67,96.00){\makebox(0,0)[cc]{= 0}}
\put(9.33,110.67){\makebox(0,0)[cc]{1}}
\put(37.33,110.67){\makebox(0,0)[cc]{2}}
\put(37.00,81.67){\makebox(0,0)[cc]{3}}
\put(9.33,81.67){\makebox(0,0)[cc]{4}}
\put(61.67,110.67){\makebox(0,0)[cc]{1}}
\put(91.00,110.67){\makebox(0,0)[cc]{2}}
\put(91.00,81.33){\makebox(0,0)[cc]{3}}
\put(61.67,81.67){\makebox(0,0)[cc]{4}}
\put(117.33,110.67){\makebox(0,0)[cc]{1}}
\put(148.67,110.33){\makebox(0,0)[cc]{2}}
\put(148.67,81.00){\makebox(0,0)[cc]{3}}
\put(117.33,81.00){\makebox(0,0)[cc]{4}}
\end{picture}

\vspace{-7.5cm}

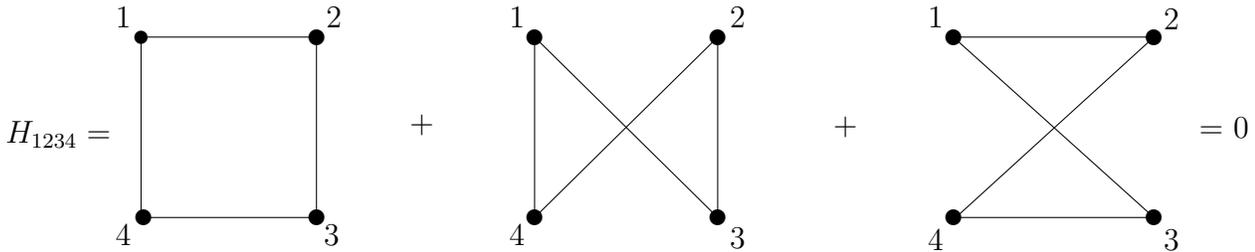
\centerline{Fig.1}

\vspace{0.5cm}

The solution $h(a)=1/a$ to this equation may be viewed as expressing
Wick's theorem for the (two-dimensional) bosonic  current
$J(a)=\d\phi(a)$ in the four-point correlation function
 \\ $\left<J(a_1)J(a_2)J(a_3)J(a_4)\right>\equiv C_{1234}$.
Wick's theorem here states that
$C_{1234}=C_{12}C_{34}-C_{13}C_{24}+C_{14}C_{23}$, where $C_{ij}$ denotes
a pair correlator. Now in a fermionic representation of the current,
$J(a)=\bar\psi(a)\psi(a)$, Wick's theorem gives us
that $C_{1234}=C_{12}C_{34}-C_{13}C_{24}+ C_{14}C_{23} +H_{1234}$,
where the first three terms are depicted in Fig.2 and the
diagrams for the fourth term are just those in Fig.1, where
$h(a_{ij})$ plays the role of the fermionic propagator.

\special{em:linewidth 0.4pt}
\unitlength 1.00mm
\linethickness{0.4pt}
\begin{picture}(149.00,111.33)
\put(65.34,104.34){\circle*{2.00}}
\put(65.34,77.34){\circle*{2.00}}
\put(90.00,77.34){\circle*{2.00}}
\put(90.00,104.34){\circle*{2.00}}
\put(90.00,104.34){\vector(1,3){0.2}}
\emline{65.34}{77.00}{1}{67.60}{77.98}{2}
\emline{67.60}{77.98}{3}{69.77}{79.05}{4}
\emline{69.77}{79.05}{5}{71.84}{80.20}{6}
\emline{71.84}{80.20}{7}{73.81}{81.45}{8}
\emline{73.81}{81.45}{9}{75.68}{82.79}{10}
\emline{75.68}{82.79}{11}{77.45}{84.22}{12}
\emline{77.45}{84.22}{13}{79.13}{85.73}{14}
\emline{79.13}{85.73}{15}{80.70}{87.34}{16}
\emline{80.70}{87.34}{17}{82.17}{89.03}{18}
\emline{82.17}{89.03}{19}{83.55}{90.81}{20}
\emline{83.55}{90.81}{21}{84.82}{92.68}{22}
\emline{84.82}{92.68}{23}{86.00}{94.65}{24}
\emline{86.00}{94.65}{25}{87.08}{96.70}{26}
\emline{87.08}{96.70}{27}{88.05}{98.84}{28}
\emline{88.05}{98.84}{29}{88.93}{101.07}{30}
\emline{88.93}{101.07}{31}{90.00}{104.34}{32}
\put(90.00,104.34){\vector(2,1){0.2}}
\emline{65.34}{77.34}{33}{66.36}{79.65}{34}
\emline{66.36}{79.65}{35}{67.45}{81.88}{36}
\emline{67.45}{81.88}{37}{68.63}{84.03}{38}
\emline{68.63}{84.03}{39}{69.88}{86.10}{40}
\emline{69.88}{86.10}{41}{71.21}{88.09}{42}
\emline{71.21}{88.09}{43}{72.61}{90.01}{44}
\emline{72.61}{90.01}{45}{74.09}{91.84}{46}
\emline{74.09}{91.84}{47}{75.65}{93.60}{48}
\emline{75.65}{93.60}{49}{77.28}{95.27}{50}
\emline{77.28}{95.27}{51}{79.00}{96.87}{52}
\emline{79.00}{96.87}{53}{80.78}{98.38}{54}
\emline{80.78}{98.38}{55}{82.65}{99.82}{56}
\emline{82.65}{99.82}{57}{84.59}{101.18}{58}
\emline{84.59}{101.18}{59}{86.61}{102.46}{60}
\emline{86.61}{102.46}{61}{90.00}{104.34}{62}
\put(90.00,77.00){\vector(3,-1){0.2}}
\emline{65.34}{104.34}{63}{65.83}{102.04}{64}
\emline{65.83}{102.04}{65}{66.45}{99.82}{66}
\emline{66.45}{99.82}{67}{67.20}{97.69}{68}
\emline{67.20}{97.69}{69}{68.06}{95.65}{70}
\emline{68.06}{95.65}{71}{69.05}{93.69}{72}
\emline{69.05}{93.69}{73}{70.16}{91.82}{74}
\emline{70.16}{91.82}{75}{71.40}{90.03}{76}
\emline{71.40}{90.03}{77}{72.75}{88.32}{78}
\emline{72.75}{88.32}{79}{74.23}{86.71}{80}
\emline{74.23}{86.71}{81}{75.84}{85.17}{82}
\emline{75.84}{85.17}{83}{77.56}{83.72}{84}
\emline{77.56}{83.72}{85}{79.41}{82.36}{86}
\emline{79.41}{82.36}{87}{81.38}{81.08}{88}
\emline{81.38}{81.08}{89}{83.48}{79.89}{90}
\emline{83.48}{79.89}{91}{85.70}{78.78}{92}
\emline{85.70}{78.78}{93}{90.00}{77.00}{94}
\put(90.00,77.00){\vector(1,-3){0.2}}
\emline{65.34}{104.34}{95}{67.60}{103.43}{96}
\emline{67.60}{103.43}{97}{69.77}{102.43}{98}
\emline{69.77}{102.43}{99}{71.84}{101.33}{100}
\emline{71.84}{101.33}{101}{73.81}{100.13}{102}
\emline{73.81}{100.13}{103}{75.68}{98.83}{104}
\emline{75.68}{98.83}{105}{77.45}{97.43}{106}
\emline{77.45}{97.43}{107}{79.13}{95.93}{108}
\emline{79.13}{95.93}{109}{80.70}{94.33}{110}
\emline{80.70}{94.33}{111}{82.17}{92.64}{112}
\emline{82.17}{92.64}{113}{83.55}{90.84}{114}
\emline{83.55}{90.84}{115}{84.82}{88.95}{116}
\emline{84.82}{88.95}{117}{86.00}{86.95}{118}
\emline{86.00}{86.95}{119}{87.08}{84.86}{120}
\emline{87.08}{84.86}{121}{88.05}{82.66}{122}
\emline{88.05}{82.66}{123}{88.93}{80.37}{124}
\emline{88.93}{80.37}{125}{90.00}{77.00}{126}
\put(10.66,104.67){\circle*{2.00}}
\put(37.66,104.67){\circle*{2.00}}
\put(37.66,77.33){\circle*{2.00}}
\put(10.66,77.33){\circle*{2.00}}
\put(37.66,104.67){\vector(2,-1){0.2}}
\emline{10.66}{104.67}{127}{12.79}{105.69}{128}
\emline{12.79}{105.69}{129}{14.93}{106.52}{130}
\emline{14.93}{106.52}{131}{17.10}{107.17}{132}
\emline{17.10}{107.17}{133}{19.29}{107.63}{134}
\emline{19.29}{107.63}{135}{21.51}{107.91}{136}
\emline{21.51}{107.91}{137}{26.01}{107.91}{138}
\emline{26.01}{107.91}{139}{28.29}{107.63}{140}
\emline{28.29}{107.63}{141}{30.60}{107.17}{142}
\emline{30.60}{107.17}{143}{32.93}{106.52}{144}
\emline{32.93}{106.52}{145}{35.29}{105.69}{146}
\emline{35.29}{105.69}{147}{37.66}{104.67}{148}
\put(37.66,104.67){\vector(2,1){0.2}}
\emline{10.66}{104.67}{149}{12.84}{103.42}{150}
\emline{12.84}{103.42}{151}{15.01}{102.38}{152}
\emline{15.01}{102.38}{153}{17.17}{101.56}{154}
\emline{17.17}{101.56}{155}{19.32}{100.94}{156}
\emline{19.32}{100.94}{157}{21.45}{100.54}{158}
\emline{21.45}{100.54}{159}{23.57}{100.35}{160}
\emline{23.57}{100.35}{161}{25.68}{100.37}{162}
\emline{25.68}{100.37}{163}{27.77}{100.61}{164}
\emline{27.77}{100.61}{165}{29.86}{101.05}{166}
\emline{29.86}{101.05}{167}{31.93}{101.71}{168}
\emline{31.93}{101.71}{169}{33.99}{102.57}{170}
\emline{33.99}{102.57}{171}{37.66}{104.67}{172}
\put(37.66,77.00){\vector(2,1){0.2}}
\emline{11.00}{77.33}{173}{12.89}{76.08}{174}
\emline{12.89}{76.08}{175}{14.82}{75.04}{176}
\emline{14.82}{75.04}{177}{16.77}{74.20}{178}
\emline{16.77}{74.20}{179}{18.76}{73.58}{180}
\emline{18.76}{73.58}{181}{20.78}{73.16}{182}
\emline{20.78}{73.16}{183}{22.84}{72.94}{184}
\emline{22.84}{72.94}{185}{24.92}{72.94}{186}
\emline{24.92}{72.94}{187}{27.04}{73.14}{188}
\emline{27.04}{73.14}{189}{29.19}{73.55}{190}
\emline{29.19}{73.55}{191}{31.38}{74.17}{192}
\emline{31.38}{74.17}{193}{33.59}{74.99}{194}
\emline{33.59}{74.99}{195}{37.66}{77.00}{196}
\put(37.66,77.33){\vector(4,-3){0.2}}
\emline{10.66}{77.33}{197}{12.78}{78.64}{198}
\emline{12.78}{78.64}{199}{14.88}{79.73}{200}
\emline{14.88}{79.73}{201}{16.98}{80.61}{202}
\emline{16.98}{80.61}{203}{19.06}{81.28}{204}
\emline{19.06}{81.28}{205}{21.13}{81.73}{206}
\emline{21.13}{81.73}{207}{23.18}{81.96}{208}
\emline{23.18}{81.96}{209}{25.23}{81.98}{210}
\emline{25.23}{81.98}{211}{27.27}{81.79}{212}
\emline{27.27}{81.79}{213}{29.29}{81.38}{214}
\emline{29.29}{81.38}{215}{31.30}{80.76}{216}
\emline{31.30}{80.76}{217}{33.30}{79.92}{218}
\emline{33.30}{79.92}{219}{35.29}{78.87}{220}
\emline{35.29}{78.87}{221}{37.66}{77.33}{222}
\put(116.67,104.67){\circle*{2.00}}
\put(142.33,104.33){\circle*{2.00}}
\put(142.33,77.00){\circle*{2.00}}
\put(116.67,77.00){\circle*{2.00}}
\put(116.67,104.67){\vector(1,2){0.2}}
\emline{116.67}{77.00}{223}{115.58}{79.16}{224}
\emline{115.58}{79.16}{225}{114.68}{81.33}{226}
\emline{114.68}{81.33}{227}{113.98}{83.51}{228}
\emline{113.98}{83.51}{229}{113.46}{85.71}{230}
\emline{113.46}{85.71}{231}{113.14}{87.91}{232}
\emline{113.14}{87.91}{233}{113.00}{90.14}{234}
\emline{113.00}{90.14}{235}{113.06}{92.37}{236}
\emline{113.06}{92.37}{237}{113.31}{94.62}{238}
\emline{113.31}{94.62}{239}{113.75}{96.88}{240}
\emline{113.75}{96.88}{241}{114.38}{99.16}{242}
\emline{114.38}{99.16}{243}{115.20}{101.44}{244}
\emline{115.20}{101.44}{245}{116.67}{104.67}{246}
\put(116.67,77.00){\vector(-1,-2){0.2}}
\emline{116.67}{104.67}{247}{117.82}{102.53}{248}
\emline{117.82}{102.53}{249}{118.78}{100.39}{250}
\emline{118.78}{100.39}{251}{119.54}{98.24}{252}
\emline{119.54}{98.24}{253}{120.11}{96.09}{254}
\emline{120.11}{96.09}{255}{120.48}{93.94}{256}
\emline{120.48}{93.94}{257}{120.65}{91.78}{258}
\emline{120.65}{91.78}{259}{120.63}{89.62}{260}
\emline{120.63}{89.62}{261}{120.42}{87.45}{262}
\emline{120.42}{87.45}{263}{120.01}{85.28}{264}
\emline{120.01}{85.28}{265}{119.40}{83.11}{266}
\emline{119.40}{83.11}{267}{118.60}{80.93}{268}
\emline{118.60}{80.93}{269}{116.67}{77.00}{270}
\put(142.33,76.67){\vector(1,-2){0.2}}
\emline{142.33}{104.33}{271}{141.34}{102.12}{272}
\emline{141.34}{102.12}{273}{140.53}{99.91}{274}
\emline{140.53}{99.91}{275}{139.89}{97.70}{276}
\emline{139.89}{97.70}{277}{139.42}{95.48}{278}
\emline{139.42}{95.48}{279}{139.12}{93.26}{280}
\emline{139.12}{93.26}{281}{139.00}{91.03}{282}
\emline{139.00}{91.03}{283}{139.06}{88.80}{284}
\emline{139.06}{88.80}{285}{139.28}{86.56}{286}
\emline{139.28}{86.56}{287}{139.68}{84.32}{288}
\emline{139.68}{84.32}{289}{140.25}{82.08}{290}
\emline{140.25}{82.08}{291}{141.00}{79.83}{292}
\emline{141.00}{79.83}{293}{142.33}{76.67}{294}
\put(142.33,76.67){\vector(-1,-3){0.2}}
\emline{142.33}{104.33}{295}{143.32}{102.22}{296}
\emline{143.32}{102.22}{297}{144.13}{100.09}{298}
\emline{144.13}{100.09}{299}{144.78}{97.94}{300}
\emline{144.78}{97.94}{301}{145.25}{95.77}{302}
\emline{145.25}{95.77}{303}{145.54}{93.58}{304}
\emline{145.54}{93.58}{305}{145.66}{91.36}{306}
\emline{145.66}{91.36}{307}{145.61}{89.12}{308}
\emline{145.61}{89.12}{309}{145.38}{86.86}{310}
\emline{145.38}{86.86}{311}{144.98}{84.58}{312}
\emline{144.98}{84.58}{313}{144.41}{82.28}{314}
\emline{144.41}{82.28}{315}{143.67}{79.96}{316}
\emline{143.67}{79.96}{317}{142.33}{76.67}{318}
\put(52.33,91.33){\makebox(0,0)[cc]{$-$}}
\put(102.33,91.33){\makebox(0,0)[cc]{$+$}}
\put(62.67,107.00){\makebox(0,0)[cc]{1}}
\put(93.00,107.00){\makebox(0,0)[cc]{2}}
\put(92.67,74.34){\makebox(0,0)[cc]{3}}
\put(62.67,74.34){\makebox(0,0)[cc]{4}}
\put(8.33,107.00){\makebox(0,0)[cc]{1}}
\put(40.33,107.00){\makebox(0,0)[cc]{2}}
\put(40.33,74.33){\makebox(0,0)[cc]{3}}
\put(8.33,74.33){\makebox(0,0)[cc]{4}}
\put(114.00,107.00){\makebox(0,0)[cc]{1}}
\put(145.33,106.67){\makebox(0,0)[cc]{2}}
\put(145.33,74.33){\makebox(0,0)[cc]{3}}
\put(114.00,74.33){\makebox(0,0)[cc]{4}}
\end{picture}

\vspace{-6cm}

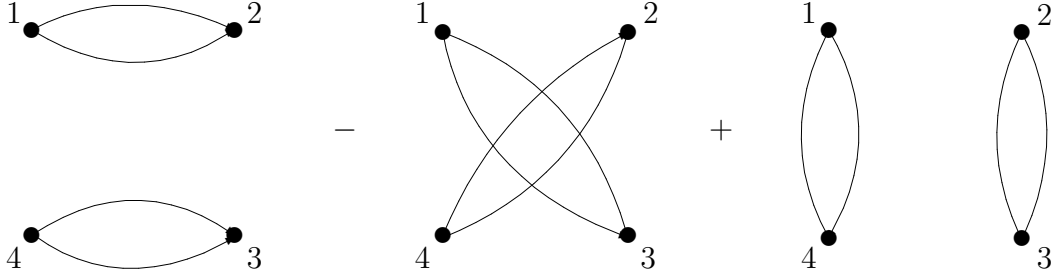
\centerline{Fig.2}

\vspace{0.5cm}
Consistency between the bosonic and fermionic realisations implies
eq.(\ref{cor}), i.e. $H_{1234}=0$. This consistency certainly holds on a
plane, where $\displaystyle{h^{plane}(a_{ij})={\sqrt{da_ida_j}\over
a_{ij}}}$. It also holds on a cylinder, where $\displaystyle{h^{cyl}
(a_{ij})={R\sqrt{da_ida_j}\over \sinh Ra_{ij}}}$.
Clearly we may obtain $\displaystyle{h^{plane}}$ from $\displaystyle{h^{cyl}}$
by taking the limit $R\rightarrow 0$. Indeed, we may also obtain
$\displaystyle{h^{cyl}}$ from $\displaystyle{h^{plane}}$ (for any $R$) by a
conformal transformation of the plane to the cylinder, since
\be
{\sqrt{da_ida_j}\over 2\sinh a_{ij}}=
{\sqrt{da_ida_j}\over e^{a_{ij}}-e^{a_{ji}}}=
{\sqrt{de^{2a_i}de^{2a_j}}\over e^{2a_i}-e^{2a_j}}.
\ee

\paragraph{5.}
It remains to relate our solution of the WDVV equations with a spectral
curve and ultimately an associated integrable system.
The perturbative prepotential is typically associated
with the spectral curve of the form $w=2P_N(\lambda)$
for pure $SU(N)$ SUSY YM theories, and  with
$w=2P_N(\lambda)/ \sqrt{Q_{N_f}(\lambda)}$
for the theory that includes $N_f$ matter hypermultiplets
(see \cite{wdvvb} for details). For the five
dimensional case of interest
we have in these formulas \cite{intd,wdvvb,theisen,5d,5dp}
$P_N(\lambda)=\prod(\lambda-e^{2a_i})$ and $Q_{N_f}=\prod(\lambda-
e^{2m_\alpha})$, where $m_\alpha$ are the hypermultiplet masses 
and $\sum\alpha_i=0$ in the $SU(N)$ case. The relevant Seiberg-Witten
differential $dS$ is given by
\be\label{dS}
dS={1\over 2}\log\lambda d\log w
\ee
In order to reproduce the prepotential
(\ref{pair}), (\ref{dli3}) discussed in the previous
sections, we stay with the same $dS$ and the same form
of the spectral
curve now choosing\footnote{To simplify formulas, hereafter we put $R=2$.}.
\be\label{pt}
w=(-)^N2{P_N(\lambda)\over P_N(-\lambda)},\ \ \ \
P_N(\lambda)=\prod_{i=1}^N
(\lambda - e^{2a_i}),\ \ \ \ \sum_{i=1}^N a_i=0
\ee
One can readily check that this leads to the correct result
in several ways. Perhaps the simplest way is simply to note that this curve
may be obtained from the curve for  YM with fundamental matter
upon choosing $N_f=2N$ with masses pairwise
coinciding and identification of these masses with $a_i+i{\pi\over 2}$.
Then the result for the prepotential obtained from
(3.37) of \cite{5dp} coincides exactly  with (\ref{pair}), (\ref{dli3}).

A more immediate way to check the curve (\ref{pt}) is to follow the
line of reasoning of \cite{wdvvb,5dp} and use 
``the residue formula" \cite{wdvvb} which takes the form
\be\label{resfor}
{\cal F}_{ijk}=
\stackreb{d\log w= 0}{\hbox{res}}
\frac{d\omega_id\omega_jd\omega_k}{d\log w d\log\lambda}.
\ee
Here $d\omega_i$ are canonical holomorphic differentials
and $dS$ is fixed to be of the form (\ref{dS}).
The derivatives of $dS$ with respect to moduli $a_i$ give the set of
\lq\lq holomorphic"
differentials on the puncture sphere (\ref{pt}), i.e.
those having simple poles only at the marked points $\lambda_i=e^{2a_i}$
(cf. with s.3.3 of \cite{5dp}). We obtain (for $i=1,\ldots N-1$)
\be
d\omega_i={\lambda_{iN} d\lambda\over 2(\lambda -\lambda_i)(\lambda
-\lambda_N)}+ {\lambda_{iN} d\lambda\over 2(\lambda +\lambda_i)(\lambda
+\lambda_N)}= \lambda_{iN}
\frac{( \lambda\sp2+\lambda_i \lambda_N)}{(\lambda\sp2-\lambda_i \sp2)(
\lambda\sp2-\lambda_N\sp2)}d\lambda,
\ee
and we have set $\lambda_{ij}\equiv\lambda_i-\lambda_j$.
Upon writing
\be
d\log w= \sum_{r=1}\sp{N}\frac{2\lambda_r d\lambda}{\lambda\sp2-\lambda_r\sp2}=
\frac{\sum_{r=1}\sp{N}2\lambda_r\prod_{t\ne r}(\lambda\sp2-\lambda_t\sp2)}
{\prod(\lambda\sp2-\lambda_s\sp2)}
\equiv \frac{H(\lambda\sp2)}{\prod_{s=1}\sp{N}(\lambda\sp2-\lambda_s\sp2)}
\ee
the residue formula (\ref{resfor}) provides
\be\label{ressub}
{\cal F}_{ijk}=
\lambda_{iN}\lambda_{jN}\lambda_{kN}
\stackreb{H(\lambda\sp2)= 0}{\hbox{res}}
\frac{
\lambda
( \lambda\sp2+\lambda_i \lambda_N)( \lambda\sp2+\lambda_j \lambda_N)
( \lambda\sp2+\lambda_k \lambda_N)
\prod_{s=1}\sp{N-1}(\lambda\sp2-\lambda_s\sp2)
}{
H(\lambda\sp2)(\lambda\sp2-\lambda_i \sp2)(\lambda\sp2-\lambda_j \sp2)
(\lambda\sp2-\lambda_k \sp2)(\lambda\sp2-\lambda_N\sp2)\sp2}.
\ee
Because $H(\lambda_i\sp2)\ne0$ and everything is well-behaved at infinity
we may exchange calculating the residues at the zeros of $H(\lambda\sp2)$
with the residues at the poles $\lambda=\pm\lambda_i$ and so on.
Consider for example the case $i,j,k$ distinct. Then
\be\label{ressubd}
{\cal F}_{ijk}=
\lambda_{iN}\lambda_{jN}\lambda_{kN}
\stackreb{H(\lambda\sp2)= 0}{\hbox{res}}
\frac{
\lambda
( \lambda\sp2+\lambda_i \lambda_N)( \lambda\sp2+\lambda_j \lambda_N)
( \lambda\sp2+\lambda_k \lambda_N)
\prod_{s\ne i,j,k}\sp{N-1}(\lambda\sp2-\lambda_s\sp2)
}{
H(\lambda\sp2)(\lambda\sp2-\lambda_N\sp2)\sp2}\\
\qquad =-
\lambda_{iN}\lambda_{jN}\lambda_{kN}
\stackreb{\lambda=\pm \lambda_N}{\hbox{res}}
\frac{
\lambda
( \lambda\sp2+\lambda_i \lambda_N)( \lambda\sp2+\lambda_j \lambda_N)
( \lambda\sp2+\lambda_k \lambda_N)
\prod_{s\ne i,j,k}\sp{N-1}(\lambda\sp2-\lambda_s\sp2)
}{
H(\lambda\sp2)(\lambda\sp2-\lambda_N\sp2)\sp2}.
\ee
Upon evaluating\footnote{
Noting
$$
\frac{\lambda}{(\lambda\sp2-\lambda_N\sp2)\sp2}
=\frac{\lambda}{(\lambda-\lambda_N)\sp2 (\lambda+\lambda_N)\sp2}
=\frac{1}{4\lambda_N}
\left(\frac{1}{(\lambda-\lambda_N)\sp2}-\frac{1}{(\lambda+\lambda_N)\sp2}
 \right)
$$
we obtain residues straightforwardly using
$$
\stackreb{\lambda=\pm \lambda_N}{\hbox{res}}
\frac{\lambda}{(\lambda\sp2-\lambda_N\sp2)\sp2}G(\lambda\sp2)
=\frac{1}{4\lambda_N}
\left(\frac{d}{d\lambda}G(\lambda\sp2)|_{\lambda_N}-
      \frac{d}{d\lambda}G(\lambda\sp2)|_{-\lambda_N}\right)
=G\sp\prime(\lambda_N\sp2).
$$
} this at the pole $\lambda=\lambda_N$  we obtain a result in agreement with
the form (\ref{g}). Similar calculations of $F_{iij}$ and $F_{iii}$
are also in agreement, verifying that we have the correct form of
the perturbative spectral curve.

The final step is going from the perturbative spectral curve to
the full spectral curve. Exactly at this step one identifies the
relevant integrable system. As conjectured in \cite{intd} the integrable
system relevant here is the elliptic Ruijsenaars model.
The details of this step will be provided elsewhere.

\paragraph{6.} The connections between integrable systems, SW theory
and the generalized WDVV equations are striking but still await a
fuller understanding. This Letter has produced another important example in this
circle of ideas.
Indeed we have turned the usual line of reasoning around
and rather than starting with a SW theory and verifying that its 
perturbative prepotential satisfies the generalized WDVV equations we
have found a new class of solutions for the generalized WDVV 
equations of a given functional form and associated to these
perturbative prepotentials for a SW theory. The theory found may be
interpreted in terms of 
a compactified  five-dimensional theory. The resulting $N=4$ four-dimensional
SUSY is broken here by non-trivial boundary conditions on half of
the fields and by varying the radius of the theory we may
interpolate between $N=4$ and $N=2$ theories. A spectral curve has been
found for this data and an integrable  system associated with the curve.
Showing here that we may reverse this  circle of ideas serves to strengthen
the need for a better  understanding of the links between
integrability and field theory.

A.Mar., A.Mir. and A.Mor. are grateful for the hospitality
of University of Edinburgh.
A.Mir. and H.W.B. also acknowledge the
Royal Society for support under a joint project.

This work was partly supported by the RFBR grants
98-01-00344 (A.Mar.), 98-01-00328 (A.Mir.) and 98-02-16575 (A.Mor.),
INTAS grants 96-518 (A.Mar.) and 96-482 (A.Mir.),
the Russian President's grant 96-15-96939 (A.Mor.) and
the program for support of the scientific schools 96-15-96798
(A.Mir.).

\end{document}